\documentstyle[aps,epsfig]{revtex}
\begin{document}
\draft
\title{Disoriented  chiral  condensate in presence of dissipation
and noise}
\author{\bf A. K. Chaudhuri \cite{byline}}
\address{
Variable Energy Cyclotron Centre\\
1-AF, Bidhan Nagar, Calcutta- 700 064}
\maketitle
\begin{abstract}
We  have investigated the phase transition and disoriented chiral
condensate domain formation in linear sigma  model.  Solving  the
Langevin  equation  for  the linear $\sigma$ model, we have shown
that for zero mass pions  the  fields  undergo  phase  transition
above a certain temperature ($T_c$). For finite mass pions, there
is  no  phase transition. It was also shown that when the fields,
thermalised  at  temperature  above   $T_c$   are   cooled   down
sufficiently  rapidly,  disoriented chiral condensate domains are
formed quite late in the evolution, if the pions are  assumed  to
be  mass  less. For massive pions, no large DCC domain is formed.
\end{abstract}
\section{introduction}

Equilibrium  high temperature QCD manifest chiral symmetry if the
quarks are assumed to be  massless.  At  a  critical  temperature
$T_c$,  chiral  symmetry is spontaneously broken by the formation
of a scalar $<\bar{q}q>$ condensate. In recent  years  there  has
been   considerable  excitement  about  the  possibility  of  the
formation  of  disoriented  chiral   condensate   (DCC)   domains
\cite{ra93a,ra93b,ra94,bj92,ko92,bj93,bj97}.  The  basic  idea is
simple.  In  hadron-hadron  or  in  heavy   ion   collisions,   a
macroscopic  region of space-time may be created within which the
chiral order parameter is not oriented in the same  direction  in
the  internal  $O(4)=SU(2)  \times  SU(2)$  space as the ordinary
vacuum. In heavy ion collisions some  region  can  thermalise  at
temperature  greater than $T_c$, the critical temperature for the
chiral symmetry restoration. In that region the quark  condensate
$<\bar{q}q>$  vanishes. Once the system cools below $T_c$, chiral
symmetry is spontaneously broken and the system  evolve  back  to
the  true  ground state where the condensate $<\bar{q}q> \neq 0$.
Rajagopal and Wilczek \cite{ra93a,ra93b,ra94} argued that in  the
non-linear  environment  (as  will  be  the  case  in  heavy  ion
collisions) as the temperature  drops  below  $T_c$,  the  chiral
symmetry  may  begin  to break by developing domains in which the
chiral field  is  misaligned  from  its  true  vacuum  value.  In
hadronic  collisions,  the space-time picture of DCC formation is
different    \cite{bj92,ko92,bj93,bj97}.     In     hadron-hadron
collisions, most of the models of particle production, stringy or
partonic,  put the bulk of the space-time activity near the light
cone. The flow of produced quanta are concentrated  in  a  rather
thin  shell  expanding  from  the collision point at the speed of
light. As the interior is  separated  from  the  true  vacuum  by
rapidly  expanding  hot  shell  of partons, well inside it, quark
condensate may  be  chirally  rotated  from  its  usual  (vacuum)
direction.  This  is  the  so  called  Baked-Alaska  model of DCC
production. The misaligned condensate has the same quark  content
and  quantum  numbers  as  do  pions and essentially constitute a
classical pion field. The system will finally relaxes to the true
vacuum and in the process can emit coherent pions. Possibility of
producing classical pion fields in heavy ion collisions had  been
discussed   earlier  by  Anslem  \cite{an91}.  DCC  formation  in
hadronic or in heavy ion collisions can lead to  the  spectacular
events  that  some  portion  of the detector will be dominated by
charged pions or by neutral pions only. In contrast, in a general
event all the three pions ($\pi^+$,$\pi^-$ and $\pi^0$)  will  be
equally  well  produced. This may then be the natural explanation
of the so called Centauro events \cite{la80}.

DCC  phenomena may be easily detected in heavy ion collisions, if
the collisions  results  in  a  single  large  DCC  domain.  Then
comparing the measured spectra of neutral and charged pions on an
event-by  event  basis,  one  could  have  identified  the events
dominated by DCC formation. The probability of a domain to  yield
a particular fraction (f) of neutral pions is $P(f)=1/2\sqrt{f}$,
provided   all   the  isospin  orientations  are  equally  likely
\cite{bj92,an91,bl92,and81}.  In  contrast,  a  typical  hadronic
reaction without DCC, will produce a binomial distribution of $f$
peaked  at  the  isospin symmetric value of 1/3. Sizes of domains
are then of great importance. A collision resulting in  a  single
large  domain  can  be easily identified from the ratio f. On the
otherhand, if the collision results in a number of small randomly
oriented  domains,  the  result  will   again   be   a   binomial
distribution, typical of a hadronic collision.

After  the  conjecture  about  DCC  formation  by  Rajagopal  and
Wilczek\cite{ra93a,ra93b,ra94}    and     also     by     Bjorken
\cite{bj92,ko92}    several   authors   studied   the   phenomena
\cite{rand,gavi,asak,ka97,cho99,cs99,kr99}.  Microscopic  physics
governing  DCC phenomena is not well known.It is in the regime of
non-perturbative QCD as well as nonlinear phenomena,  theoretical
understanding  of  both  of which are limited. One thus uses some
effective field theory like linear $\sigma$  model  with  various
approximations  to  simulate  the chiral phase transition. Linear
$\sigma$ model usually with Hartree  type  of  approximation  has
been    used    extensively    to    simulate    DCC    phenomena
\cite{rand,asak,gavi}. In the linear sigma model  chiral  degrees
of   freedom   are   described   by   the  the  real  O(4)  field
$\Phi=(\sigma,\roarrow{\pi})$. Because of the isomorphism between
the groups $O(4)$ and $SU(2) \times SU(2)$, the later  being  the
appropriate  group  for  two  flavor  QCD, linear sigma model can
effectively model the low  energy  dynamic  of  QCD  \cite{bo96}.
Rajagopal  and  Wilczek  \cite{ra93a,ra93b,ra94}  used  of quench
initial condition to simulate the  non-equilibrium  evolution  of
the  chirally  symmetric  field.  Quench scenario assume that the
effective potential governing the evolution of long  wave  length
modes  immediately  after  the phase transition at $T_c$ turns to
classical one at zero temperature. It can happen only in case  of
very  rapid  cooling  and expansion of the fireball. In heavy ion
collisions quench like initial conditions are unlikely. Gavin and
Mueller\cite{gavi} considered the annealing  scenario,  in  which
the condensate evolve from $<\phi> \sim 0$ to $<\phi> \sim f_\pi$
slowly   (as   in  the  standard  annealing)  but  in  which  the
non-equilibrium conditions  which  allow  amplification  of  long
wavelength modes are maintained nevertheless. Explicit simulation
with  linear sigma model, indicate that DCC depends critically on
the  initial  field  configurations.  With  quench  like  initial
condition  DCC  domains  of  4-5 fm in size can form \cite{asak}.
Initial conditions other than quench lead to much smaller  domain
size.

Very  recently, effect of external media on possible DCC is being
investigated
\cite{bi97,gr97,xu99,ch99a,ch99b,ch99c,ch00a,ch00b,ri98}. Indeed,
in  heavy  ion  collisions,  even if some region is created where
chiral symmetry is restored,  that  region  will  be  continually
interacting  with  surrounding  medium  (mostly  pions). Biro and
Greiner \cite{bi97} using a  Langevin  equation  for  the  linear
$\sigma$  model, investigated the interplay of friction and white
noise on the evolution and stability of collective  pion  fields.
In  general  friction and noise reduces the amplification of zero
modes  (they  considered  the  zero  modes  only).  But  in  some
trajectories,   large  amplification  may  occur.  We  have  also
investigated the interplay of friction and noise on possible  DCC
phenomena     \cite{ch99a,ch99b,ch99c,ch00a,ch00b}.    Simulation
studies in 2+1 dimension indicate that with quench  like  initial
field  configuration, large disoriented chiral condensate domains
can still be formed. We have also performed a simulation  in  3+1
dimension.  It  was  seen  that  with  zero mass pions, even with
thermalised fields as  initial  condition,  DCC  domains  may  be
obtained, late in the evolution. However, as will be shown in the
present  paper, with finite mass pions, the different results are
obtained. We find no DCC domain structure  of  appreciable  size,
even late at evolution.

In  the  present  paper,  we  have  investigated  in  detail  the
properties  of  phase  transition  in  linear  sigma  model   and
subsequent  DCC  domain  formation in 3+1 dimension. The paper is
organised as follows: in  section  2,  we  briefly  describe  the
Langevin  equation  for  linear  $\sigma$ model and show that for
zero mass pions, the model undergoes 2nd order  phase  transition
at  finite temperature. For massive pions there is no exact phase
transition in the model, though the symmetry  is  restored  to  a
large  extent  at  high  temperature.  In  section  3,  growth of
disoriented chiral condensate domains will be studied. It will be
shown that while with zero pion mass, large DCC domains can form,
with fast cooling law, for massive  pions  no  such  domains  are
formed.  In  section 4, we will vindicate the results obtained in
section 3, from study of pion-pion correlation.  The  probability
distribution function for the neutral to total pion ratio will be
obtained  in  section  4.  Lastly,  in  section  5,  summary  and
conclusions will be drawn.

\section{Linear sigma model and phase transition}

The linear sigma model Lagrangian can be written as,

\begin{equation}
{\mathcal{L}} =\frac{1}{2} (\partial_\mu \Phi)^2 -
\frac{\lambda}{4}
(\Phi^2 - v^2)^2 + H_\sigma, \label{1}
\end{equation}

\noindent  where  chiral  degrees  of freedom are the O(4) fields
$\phi_a =(\sigma, \roarrow{\pi})$. In  eq.\ref{1}  $H_\sigma$  is
the explicit symmetry breaking term. This term is responsible for
finite  pion mass. The parameters of the model $\lambda$, $v$ and
$H$ can be fixed using  the  pion  decay  constant  $f_\pi$,  and
$\sigma$  and  pion  masses. With standard parameters, $f_\pi$=92
MeV, $m_\sigma$=600 MeV and $m_\pi$=140 MeV, one obtain,

$$\lambda=\frac{m_\sigma^2-m_\pi^2}{2f^2_\pi}       \sim      20,
v^2=f^2_\pi-\frac{m^2_\pi}{\lambda}=(87     MeV)^2,     H=f_\pi
m^2_\pi= (122 MeV)^3$$

In  eq.\ref{1},  the  potential  corresponds  to zero temperature.
  At  finite  temperature  (T),  to  leading  order  in
$\lambda$,  thermal  fluctuations  $<\delta \phi^2>$ of the pions
and sigma mesons generate an  effective  Hartree  type  dynamical
mass giving rise to an effective temperature dependent potential.
In the high temperature expansion, this results in \cite{bo96},

\begin{equation}
m^2_{th} \rightarrow  \frac{\lambda}{2} T^2
\end{equation}

As  told  in  the beginning, if in heavy ion collision, a certain
region undergoes chiral symmetry restoration, that region must be
in contact with some environment or background. Exact  nature  of
the  environment is difficult to determine but presumably it will
be  consists  of  mesons  and  hadron  (pions,   nucleon   etc.).
Recognising   the   uncertainty   in  the  exact  nature  of  the
environment, we choose to represent it by a white  noise  source,
i.e.  a  heat  bath. To analyse the effect of environment or heat
bath on the possible disoriented chiral condensate, we propose to
study Langevin  equation  for  linear  $\sigma$  model.  Langevin
equation  for O(4) fields has been used previously to investigate
the interplay of noise  and  dissipation  on  the  evolution  and
stability  DCC.  Recently  it has been shown that in the $\phi^4$
theory, hard modes can be integrated out  on  a  two  loop  basis
resulting  in  a  Langevin  type  equations  for  the  soft modes
\cite{gr97,xu99},  with  dissipation  and  noise.  The  resulting
equation  is  non-local  in  time  and  quite  complicated.  If a
Markovian limit exists to this equation, then  one  may  hope  to
obtain  a simpler equation (as the present equation) which can be
used for practical purposes.

We write the  Langevin  equation  for linear $\sigma$ model as,

\begin{equation}
[\frac{\partial^2}{\partial \tau^2} +(\frac{1}{\tau}+\eta)
\frac{\partial}{\partial \tau}
-\frac{\partial^2}{\partial x^2} -\frac{\partial^2}{\partial y^2} -
\frac{1}{\tau^2} \frac{\partial^2}{\partial Y^2}
+\lambda (\Phi^2 - f^2_\pi -T^2/2)] \Phi
 = \zeta (\tau ,x,y,Y)\label{1a}
\end{equation}

\noindent  where $\tau$ is the proper time and Y is the rapidity,
the appropriate coordinates  for  heavy  ion  scattering.  To  be
consistent with fluctuation-dissipation theorem, we have included
a  dissipative  term $\eta$ in the equation. As told earlier, the
environment or the heat bath ($\zeta$) was represented as a white
noise source with zero average and  correlation  as  demanded  by
fluctuation-dissipation theorem,

\begin{mathletters}
\begin{eqnarray}
<\zeta(\tau,x,y,Y)> =&&0\\
 <\zeta_a(\tau,x,y,Y) \zeta_b(\tau^\prime,x^\prime,y^\prime,Y^\prime)>
=&& 2 T \eta \frac{1}{\tau}
\delta(\tau-\tau^\prime)   \delta(x-x^\prime)  \delta(y-y^\prime)
\delta(Y-Y^\prime)
\delta_{ab}
\end{eqnarray}
\label{1b}
\end{mathletters}

\noindent  where a,b corresponds to $\pi$ or $\sigma$ fields. The
noise  term  will  continuously  heat  the  system,   while   the
dissipative term will counteract it. Equilibrium is achieved when
the system is thermalised at the temperature dictated by the heat
bath. We note that eq.\ref{1} can not be derived formally, but as
will  be  shown  below,  it does describe the equilibrium physics
correctly. Thus at least on phenomenological level, its  use  can
be justified, and we can hope that it will describe correctly the
non-equilibrium physics , as required for the DCC phenomena.

Before  we  proceed  further  few  words  are necessary about the
applicability of Langevin equation in describing  DCC,  when  the
noise   term   contains   temperature.   DCC   is   basically   a
non-equilibrium  phenomena.  By   using   temperature,   we   are
approximating  it  as  a  equilibrium  one. This approximation is
valid  when   the   system   is   not   far   from   equilibrium.
Fluctuation-dissipation  relation is also valid for such a system
only. In such a system, it is possible to define a temperature at
each point of time, if the time scale of the constituents of  the
system  is  large  compared  to  the time scale of the collective
variable (in this case expansion of the system).

Set  of partial differential eqs. \ref{1} were solved on a $32^3$
lattice using a lattice  spacing  of  1  fm  and  using  periodic
boundary   conditions.   Solving   eqn.\ref{1}   require  initial
conditions ($\phi$ and $\dot{\phi}$). We distribute  the  initial
fields according to a random Gaussian with,

\begin{mathletters}
\begin{eqnarray}
<\sigma>=&&(1-f(r))f_\pi \\
<\pi_i>=&&0 \\
<\sigma^2>-<\sigma>^2 = <\pi_i^2>-<\pi_i>^2=   && f_\pi^2/4 f(r)\\
< \dot{\sigma}>=&& <\dot{\pi_i}>=0\\
<\dot{\sigma}^2>=<\dot{\pi}>^2=&& f_\pi^2 f(r)
\end{eqnarray}
\label{2}
\end{mathletters}

The interpolation function

\begin{equation}
f(r)=[1+exp(r-r_0)/\Gamma)]^{-1}
\end{equation}

\noindent  separates  the  central  region  from  the rest of the
system. We have  taken  $r_0$=11  fm  and  $\Gamma$=0.5  fm.  The
initial field configurations corresponds to quench like condition
\cite{ra93a,ra93b}  but  it  is  important to note that the field
configuration at  equilibrium  will  be  independent  of  initial
configuration.  The other parameter of the model is the friction.
The friction coefficient  was  assumed  to  be  $\eta=\eta_\pi  +
\eta_\sigma$,   where   $\eta_{\pi,\sigma}$   are   the  friction
coefficient of the pion and the  sigma  fields.  They  have  been
calculated by Rischke \cite{ri98},

\begin{mathletters}
\begin{eqnarray}
\eta_\pi =&& (\frac{4 \lambda f_\pi}{N})^2 \frac{m_\sigma^2}{4  \pi
m_\pi^3} \sqrt{1-\frac{4 m_\pi^2}{m_\sigma^2}}
\frac{1-exp(-m_\pi/T)}{1-exp(-m_\sigma^2/2 m_\pi T)}
\frac{1}{exp(m_\sigma^2-2m_\pi^2)/2m_\pi T)-1}\\
\eta_\sigma   =&&  (\frac{4  \lambda  f_\pi}{N})^2  \frac{N-1}{8\pi
m_sigma}
\sqrt{1-\frac{4 m_\pi^2}{m_\sigma^2}} \coth \frac{m_\sigma}{4T}
\end{eqnarray} \label{2a}
\end{mathletters}

Equilibrium physics will be independent of the exact value of the
friction  also.  Friction  just  determine  the  rate of approach
towards the equilibrium.

To  show  that  equilibrium physics is independent of the initial
field configurations, in fig.1, we have shown the  (proper)  time
evolution of the condensate value of $\sigma$ field,

\begin{equation}
<\sigma> =\frac{1}{V} \int \sigma dx dy dY
\end{equation}

\noindent  for  two different initial conditions. In one of them,
the initial configuration was  chosen  such  that  $<sigma>  \sim
f_\pi$,  (depicted in fig.1 as white symbol) and in the other the
initial configuration was  chosen  such  that  $<sigma>  \sim  0$
(depicted  in the figure by the filled symbol). We have shown the
evolution of the condensate for two fixed temperature of the heat
bath, 20 MeV and 40 MeV.  Equilibrium  configuration  is  reached
about  10  fm  of  evolution.  For  both  the  temperatures,  the
equilibrium condensate value is independent of the initial value.
It is also noted that the at higher temperature,  the  $<\sigma>$
condensate (as expected) settles to a lower value.

Equilibrium  configuration is also independent of the exact value
of the friction. In fig.2,  we  have  plotted  the  evolution  of
$\sigma$  condensate  for  two  different values of the friction;
$\eta$ as given in eq.\ref{2} and half its value. Here  again  we
perform  the  calculation  for  two different temperatures of the
heat bath, 20 and 40 MeV. It can be seen the equilibrium value of
the condensate donot depend on the exact value of  the  friction.
Friction determines the rate at which equilibrium is achieved.

Having  thus  shown  that the equilibrium value of the fields are
independent of the initial conditions or the exact value  of  the
friction,  we  now  proceed  to study the phase transition in the
model. To this end, we keep the heat bath  at  fixed  temperature
(T)  and  evolve  the fields for sufficiently long time such that
equilibrium condition is reached. The  condensate  value  of  the
$\sigma$  field  at  equilibrium  can  be considered as the order
parameter for the chiral phase transition. In the symmetry broken
phase it will have non-zero value while in the symmetric phase it
will  vanish.  We  note  that  for  symmetry  restoration   phase
transition  the  order  parameter should {\em exactly} vanish for
temperatures $\geq$ $T_c$ \cite{landau}. Small but non-zero value
of the condensate will indicate approximate symmetry restoration.

In  fig.3,  we have shown the equilibrium condensate value of the
sigma field at different temperatures. With the explicit symmetry
breaking term in \ref{1} the pions are massive. Thus  one  should
not get exact symmetry restoration phase transition in the model.
Indeed,  in  our  simulation  also, exact phase transition is not
obtained. At low temperature  condensate  value  as  expected  is
around  $f_\pi$,  and  it  decreases  with  temperature. However,
though it became very small, it donot vanish even at  very  large
temperature.  Thus  there  is  no  exact  phase transition in the
model. However very small value of $<\sigma>$ at high temperature
in comparison to its zero temperature value, indicate that  there
is partial restoration of the symmetry.

If  the  explicit symmetry breaking term is neglected, then there
is exact  chiral  symmetry  in  the  model.  In  fig.4,  we  have
presented the equilibrium condensate values of the sigma field as
a  function  of temperature for zero mass pions. Now we find that
sigma condensate exactly vanishes for temperature higher 120 MeV.
The model shows that with zero mass pions, there  is  a  symmetry
restoring phase transition at high temperature.

\section{Disoriented chiral condensate domain formation}

In  the  last  section,  we have shown that the Langevin equation
\ref{1} correctly reproduces  the  equilibrium  behavior  of  the
linear  $\sigma$ model fields. For zero mass pions, simulation of
the Langevin equations shows symmetry restoring phase  transition
at  appropriate temperature. For massive pions it was shown that,
though there is no exact  symmetry  restoring  phase  transition,
nevertheless symmetry is restored to a great extent.

Total  or  partial  chirally  symmetric phase (as is the case for
zero or finite pion mass) at high temperature will roll  back  to
symmetry  broken  phase as the system cools and temperature drops
below the transition temperature. As told earlier, it is has been
conjectured \cite{ra93a,ra93b} that during the roll down  period,
pseudo  scalar  condensate $<\bar{q} \tau \gamma_5 q>$ can assume
non vanishing values, (instead of of remaining  zero  as  in  the
ground state). In the process domain like structure with definite
isospin  orientation  may  emerge  during  the  roll down period.
Numerical simulations of linear  sigma  model  with  quench  like
initial   field   configurations  shows  domain  like  structures
\cite{asak}. It was also seen that $\pi \pi$ correlation is  also
increased \cite{asak}. However, in heavy ion collision, quench is
not  a natural initial condition. $<\phi>$ and $<\dot{\phi}>$ are
in a configuration appropriate for high temperature but  that  of
$<\phi^2>$  and  $<\dot{\phi}^2>$  are  characteristic of a lower
temperature. On  the  contrary,  thermalised  fields  are  better
suited  to  mimic  initial conditions that may arise in heavy ion
collisions. Here, $<\phi>$, $<\dot{\phi}>$ as well as  $<\phi^2>$
and  $<\dot{\phi}^2>$ are in a configuration appropriate for high
temperature.

To  see  whether  domain  like structure emerges or not with more
appropriate initial condition like  the  thermalised  fields,  we
proceed  as  follows:  using  the  Langevin  equation \ref{1}, we
thermalised the fields at temperature T=200 MeV. Next we use  the
thermalised  fields as the initial configuration and now we allow
the heat bath to cool and follow the evolution of the thermalised
fields. Evolution of the thermalised fields now  will  depend  on
the  exact  nature  of the friction, however we choose to use the
same friction as before. We assume the following cooling law  for
the heat bath,

\begin{equation}
T(t)=T_0 \frac{1}{t^n}
\label{4}
\end{equation}

\noindent  with n=1, appropriate for 3d scaling expansion, and we
call it fast cooling law. We have also used n=1/3 (slow cooling),
appropriate  for  1-dimensional  scaling   expansion   for   some
demonstrative calculations.

Assuming that the number density is proportional to the square of
the  fields  amplitude  ($N_\pi  \propto  \phi_\pi^2$),  at  each
lattice point, we calculate  the  neutral  to  total  pion  ratio
according to,

\begin{equation}
f(x,y,Y) =\frac{N_{\pi_0}}{N_{\pi_1} + N_{\pi_2} +N_{\pi_0}}
\end{equation}

Very  large or small value of the ratio, over an extended spatial
zone will be definite indication of disoriented chiral condensate
domain formation.

\subsection{DCC domains without explicit symmetry breaking term}

We  first  consider the case for zero mass pions, i.e. neglecting
the symmetry breaking term in eq.\ref{1}.  For  zero  mass  pions
there  is  a  exact  symmetry  restoring  phase  transition,  the
condensate value of the $<\sigma>$ field  vanishes  around  T=120
MeV.  In  fig.5,  evolution of the neutral to total pion ratio at
rapidity  $Y=0$,  with  fast  cooling,  is  shown.  The   initial
distribution (panel a) do not show any domain like structure with
very  high  or  low  value  of  the  ratio  (the distribution was
random).  No  domain  like   structure   is   seen   even   after
thermalisation of the fields at T=200 MeV (panel b). Large domain
like  structure  with  high/low  value of the ratio $f$ starts to
emerge after 10-15 fm of evolution and cooling. With time  domain
like  structure  grow.  It  seems  that indeed disoriented chiral
condensate domains are formed in presence of noise and  friction,
even  with  thermalised fields. In fig.6 and 7, we have shown the
same results but at different  rapidities.  At  other  rapidities
also,  similar  behaviour  is  seen. Fig.5-7, clearly demonstrate
that  with  fast  cooling  law,   multiple   disoriented   chiral
condensate domains with definite isospin orientations can form as
the  chirally  symmetric phase roll back to broken phase. However
domain formation occur quite late  in  the  evolution.  With  the
scaling  cooling law, by the time domain like structures emerges,
the system is cooled to $\sim$20 MeV. Applicability of  classical
field  theory down to that low temperature is questionable. Apart
from that it is doubtful whether in heavy ion  collision,  system
can  be  allowed  to  be cooled to this extent. Current wisdom is
that  the  hadrons  freeze-out  around  100-160  MeV.  With  this
reservation in mind, it may be said that even with thermal fields
domains of disoriented chiral condensate can form. But the domain
growth occur at late times.

To  see  whether  domain  like  formation  occur if the system is
cooled slowly, we have repeated the above  calculation  with  the
cooling  law  corresponding  to 1d scaling law. In fig.8, contour
plot of the neutral to total pion ratio at rapidity Y=0 is shown.
It can be seen that in this case there is no extendent zone  with
very  large or small $f$, indicating domain like formation is not
likely when the expansion or cooling is slower.  Similar  results
are  obtained  at  other  rapidities also. If the cooling is slow
then  the  fields  gets  enough  time  to  adjust  to  their  new
environment, consequently energy can be equipartitioned among all
the  modes.  In heavy ion collision, 1d scaling expansion is more
appropriate than 3d scaling expansion.  Present  simulation  then
indicate  that  it is unlikely that large size disoriented chiral
condensate domains will be formed in heavy ion collisions.

\subsection{DCC domains with explicit symmetry breaking term}

For  finite mass pions, there is no exact phase transition in the
model. However, symmetry is partially restored.  To  see  whether
DCC  domains  are  formed  even  when  there  is  no  exact phase
transition,  we  repeat  the  above  calculation  including   the
symmetry  breaking term. In fig.9, xy contour plot of the neutral
to total pion ratio at rapidity $Y=0$ is shown. The  cooling  was
fast.  We  donot  find any domain like configuration. With slower
cooling law also, though not shown, we donot find any  indication
of  DCC domains formation. Indeed, if domains are not formed with
fast cooling law, then it is unlikely that they  will  be  formed
with  a  slow  cooling  law. The simulation results indicate that
there is no disoriented chiral  condensate  domains  with  finite
mass pions.

\section{Pion-pion correlation}

We define a correlation function at rapidity $Y$ as \cite{asak},

\begin{equation}   C(r,\tau)  =  \frac{  \sum_{i,j}  \pi(i)  \dot
\pi(j)}{\sum_{i,j}
|\pi(i)| |\pi(j)|}
\end{equation}

\noindent  where  the sum is taken over those grid points i and j
such that the distance between i and j is r. As DCC  domains  are
not  formed,  irrespective of pion mass with slow cooling law, in
the following we will present the  correlation  results  for  the
fast  cooling  law  only.  In fig. 10, we have shown the temporal
evolution of the correlation  function  at  rapidity  $Y=0$,  for
finite  mass  pions.  Initially,  the  thermalised  pions have no
correlation length beyond the lattice spacing of 1 fm. With  time
correlation   length  increases.  But  really  large  long  range
correlation builds up quite late in the evolution, after 10-15 fm
of  evolution.  Pions  separated  by  large  distances  are  then
correlated.  The  results are in accordance with the contour plot
of the neutral to total pion ratio, showing DCC domain  formation
for zero mass pions at late times.

With massive pions, results are quite different (fig.11). In this
case  as shown earlier, there is no exact phase transition. Also,
evolution of thermalised fields do not results into large  domain
like  formation.  Similarly,  we find that there is not much long
range correlation even at late stage of the evolution.

\section{probabilty distribution of neutral to pion ratio}

If  a  single DCC domain is formed in heavy ion collision, it can
be easily detected. Probability to obtain a particular fraction

\begin{equation}
f=\frac{N_{\pi^0}}{N_{\pi^+}+N_{\pi^-}+N_{\pi^0}}
\end{equation}

\noindent   of   neutral   pion   from   a   single   domain   is
$P(f)=1/2\sqrt{f}$ \cite{ra93a,ra93b}. However,  our  simulations
indicate that if at all domains are formed, there will be quite a
few  number of domains. Naturally the resultant distribution will
not  be  $1/\sqrt{f}$  type.  In  this  section  we  obtain   the
probability distribution of neutral to total pion ratio.

We  calculate  the  number  of pions at rapidity Y by integrating
square of the amplitude of the pion fields  over  the  space-time
as,

\begin{equation}
N_\pi(Y) = \int \pi^2 \tau d\tau dx dy
\end{equation}

In  fig.12,  probability  distribution  for neutral to total pion
ratio, for zero mass pions are shown. In three  panels,  we  have
shown  the  distribution  obtained  after  10,20  and  30  fm  of
evolution. If the fields are evolved upto 10  fm,  then  the  $f$
distribution  is  sharply  peaked  around  the  isospin symmetric
average value of 1/3. At the early stage  of  the  evolution,  as
indicated  in  figs.5-7,  domain  like  structures  or long range
correlations  are  not  developed  in  individual  events.  As  a
consequence  of  that, the neutral to pion ratio is peaked around
the isospin symmetric value of 1/3. If the fields are evolved for
longer duration (20  or  30  fm),  average  of  the  distribution
remains  1/3,  but  it  gets broadened. As shown earlier, at late
stage of the evolution there is definite domain like structure in
individual events. Long range  correlations  also  develops.  Its
effect is seen as the increased width. However, increase in width
is not as dramatic as expected from the contour plot of the ratio
or  the  correlation studies. In obtaining the f-distribution, we
have integrated the pion amplitudes over the time. And  as  shown
earlier,  domain  like structure develop only at late times. Thus
in a single event also, non-DCC pions will be mixed up  with  DCC
pions.

However,  quite  different  results  are obtained if the symmetry
breaking term is included in the Langevin equation. The pions are
massive then. Probability distribution of neutral to  total  pion
ratio,  for massive pions are shown in fig.12. Whether the fields
are evolved upto 10, 20 or 30 fm,  the  distribution  is  sharply
peaked  around  the  isospin symmetric value of 1/3. We have seen
that for massive pions, no large domain like  structure  or  long
range  correlation  develop  even  at late times. The probability
distribution reflect those results.

\section{Summary and Conclusions}

In  summary,  we  have  studied the disoriented chiral condensate
formation in presence of background. We have  simulated  Langevin
equation  for  linear sigma model with a heat bath (the heat bath
represents the background) in 3+1 dimension. It  was  shown  that
the model reproduces equilibrium physics correctly. For zero mass
pions,  the  model  undergoes 2nd order phase transition at $T_c$
$\sim$ 120 MeV. Also the model (correctly) donot show exact phase
transition for finite mass pions, even  though  the  symmetry  is
restored to a great extent at high temperature. It was also shown
that,  for  zero mass pions, if the thermalised fields are cooled
down sufficiently rapidly, multiple disoriented chiral condensate
domains are  formed,  while  for  slow  cooling  no  domain  like
structure is formed. Long range correlation also develops in case
of fast cooling. However, even with sufficient fast cooling, with
thermalised  fields,  domains  or long range correlations develop
quite late in  the  evolution  (after  10-15  fm  of  evolution).
Probability  distribution  of  neutral  to total pion ratio shows
that the distribution is not $1/\sqrt{f}$ type, rather  it  is  a
Gaussian  with average at the isospin symmetric value of 1/3. The
width of the Gaussian increases, when domains are formed.

For massive pions our simulation indicate that thermalised fields
donot  evolve into domain like structure, irrespective of slow or
fast cooling law. No long range correlation  is  also  seen.  The
probability  distribution  of  neutral  to  total pion ratio also
donot show any broadening of width as  was  seen  for  zero  mass
pions.

To  conclude, simulation studies of Langevin equations for linear
sigma model  indicate  that  for  realistic  pions,  thermalised
fields donot evolve into DCC domain like structure.

\acknowledgements
The  author gratefully acknowledge the kind hospitality of Centre
for Theoretical Physics, MIT and McGill University where part  of
the work was done.

\begin{figure}
\centerline{\psfig{figure=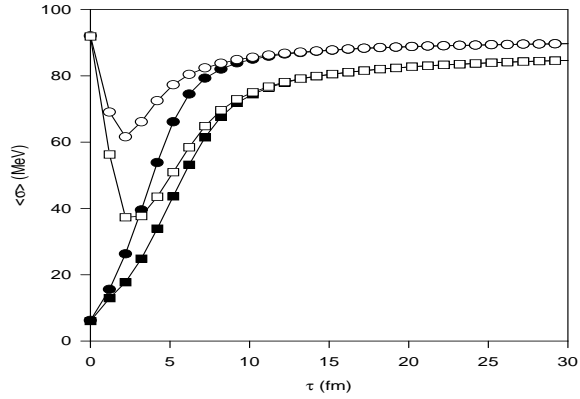,height=10cm,width=10cm}}
\caption{Evolution  of the $\sigma$ condensate with (proper) time
for two different fixed temperature 20 MeV and 40 MeV of the heat
bath.  Two  different  initial  condition  evolve  to  the   same
condensate value (see text).}
\end{figure}

\begin{figure}
\centerline{\psfig{figure=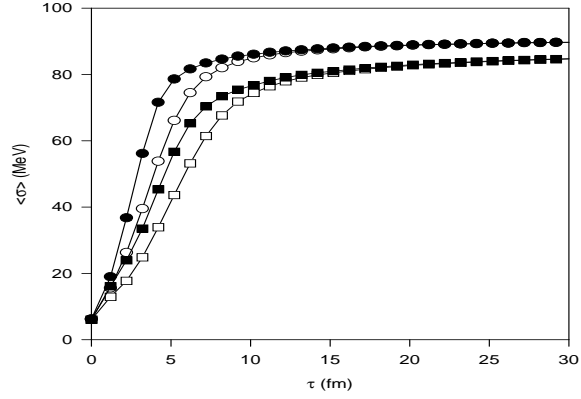,height=10cm,width=10cm}}
\caption{Evolution  of the $\sigma$ condensate with (proper) time
for two different fixed temperature 20 MeV and 40 MeV of the heat
bath, showing that equilibrium value is independent  of  friction
constant.  Two  different  friction  constant  results  into same
equilibrium value (see text).}
\end{figure}

\begin{figure}
\centerline{\psfig{figure=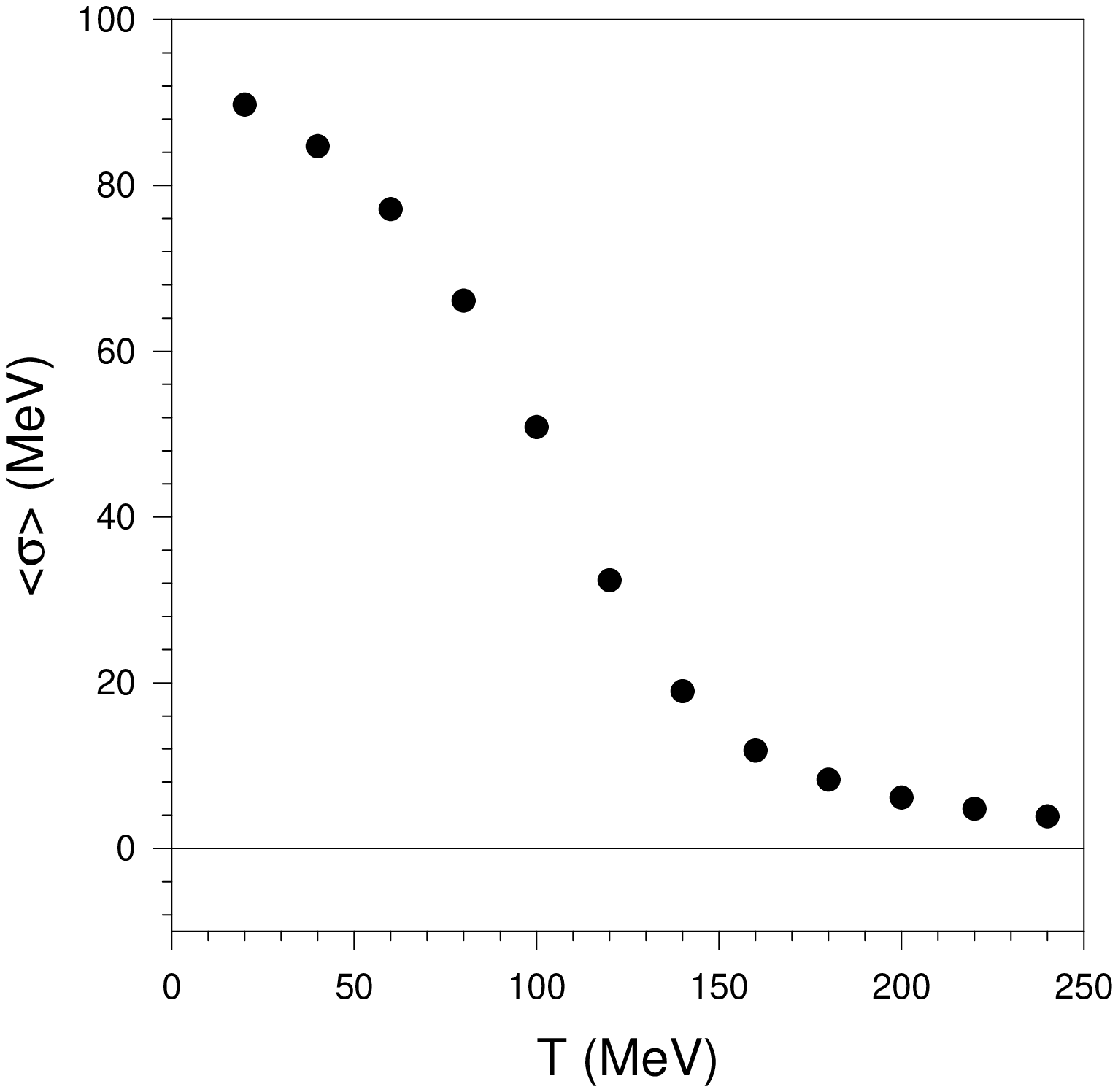,height=10cm,width=10cm}}
\caption{Equilibrium  value  of  the  $\sigma$  condensate  as  a
function of temperature. The pions are massive.}
\end{figure}

\begin{figure}
\centerline{\psfig{figure=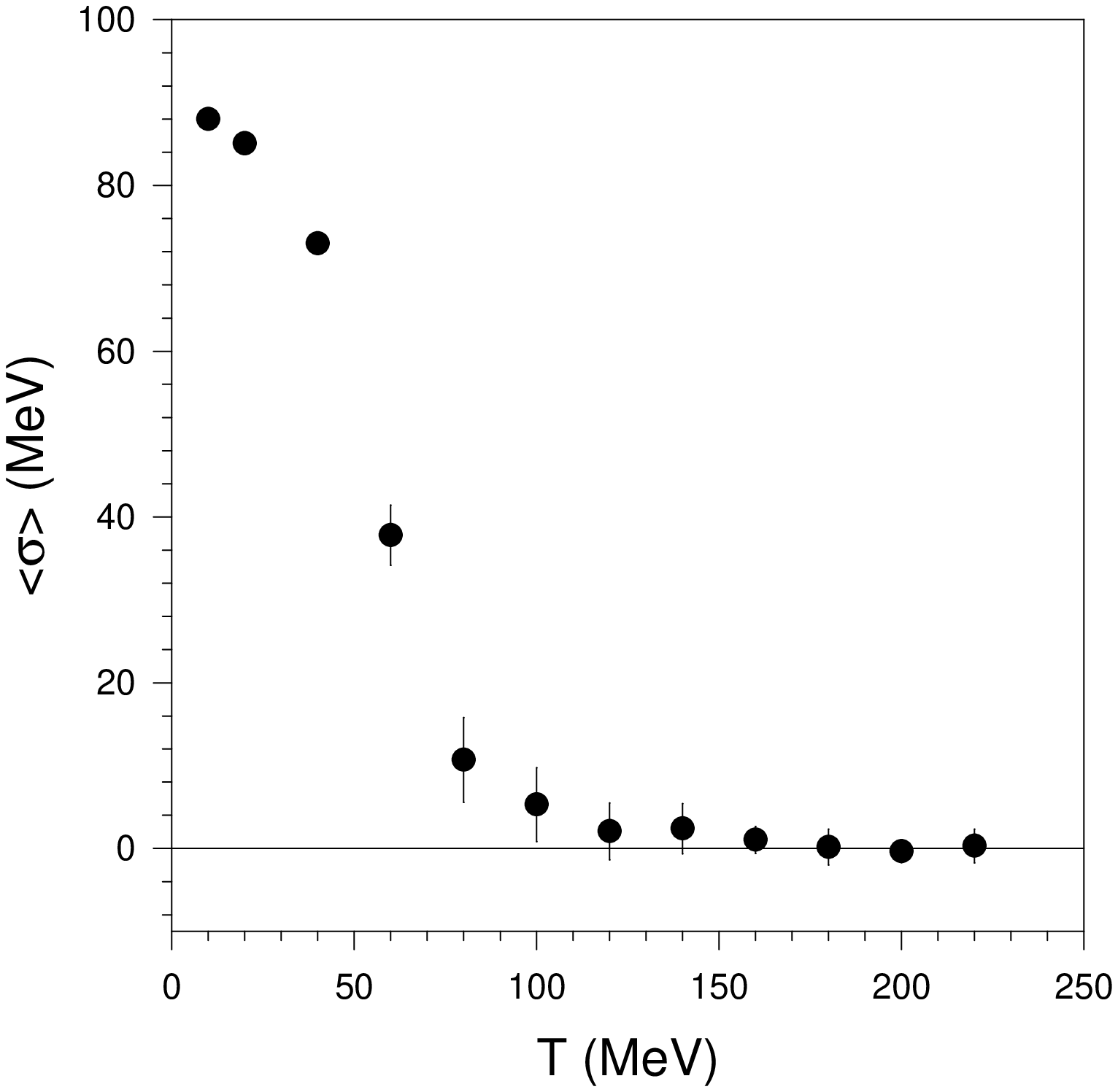,height=10cm,width=10cm}}
\caption{Equilibrium  value  of  the  $\sigma$  condensate  as  a
function of temperature. The pions are massless.}
\end{figure}

\begin{figure}
\centerline{\psfig{figure=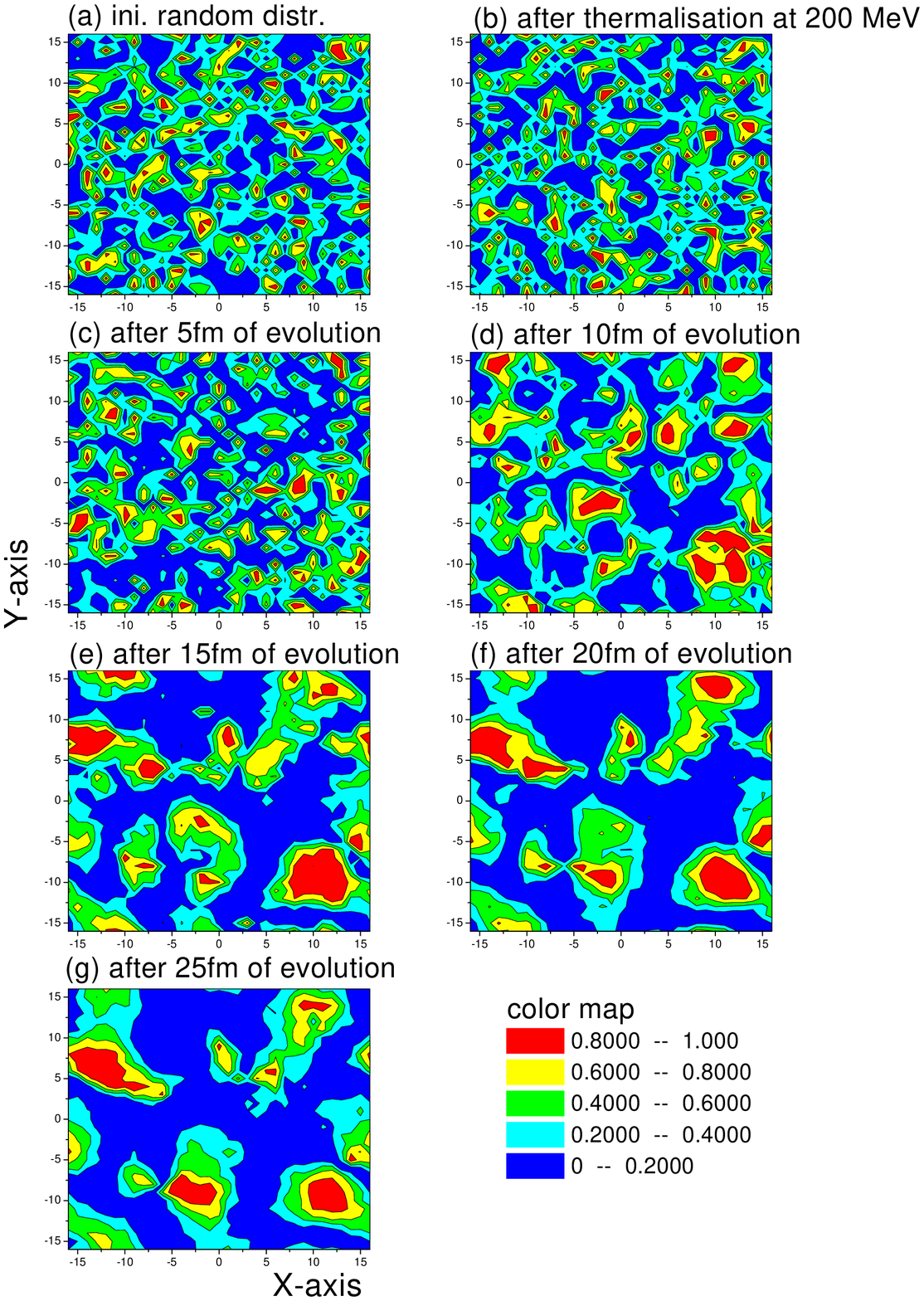,height=10cm,width=10cm}}
\caption{Contour  plot  of the neutral to total pion ratio, at
rapidity Y=0. The explicit symmetry breaking term is omitted i.e.
pions are massless. The cooling law corresponds to  fast  cooling
law.  Different  panels  shows  the  evolution  of  the  ratio at
different times. Domain like structure is evident at late times.}
\end{figure}

\begin{figure}
\centerline{\psfig{figure=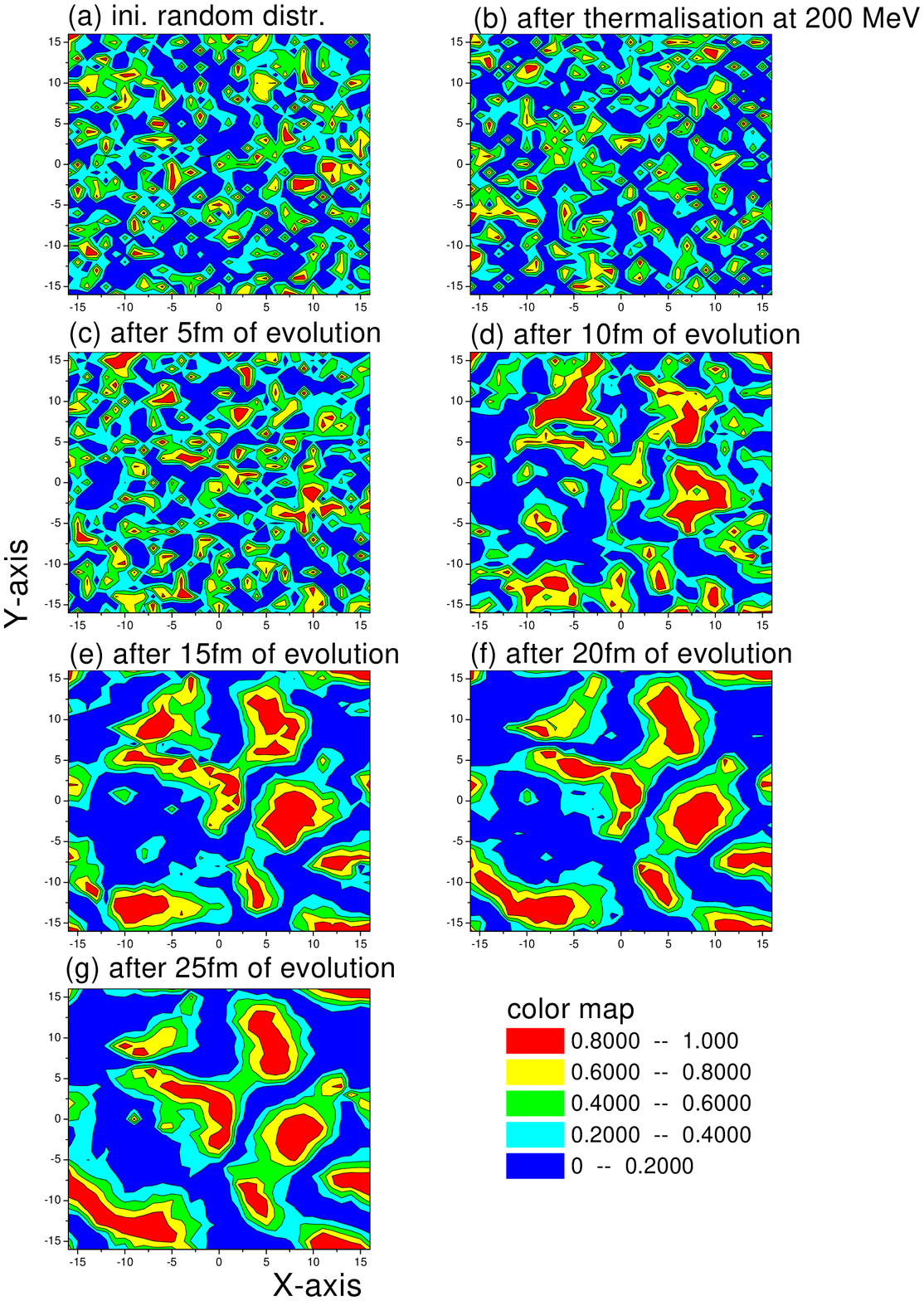,height=10cm,width=10cm}}
\caption{Same as fig.5, but for rapidity Y=-4.}
\end{figure}

\begin{figure}
\centerline{\psfig{figure=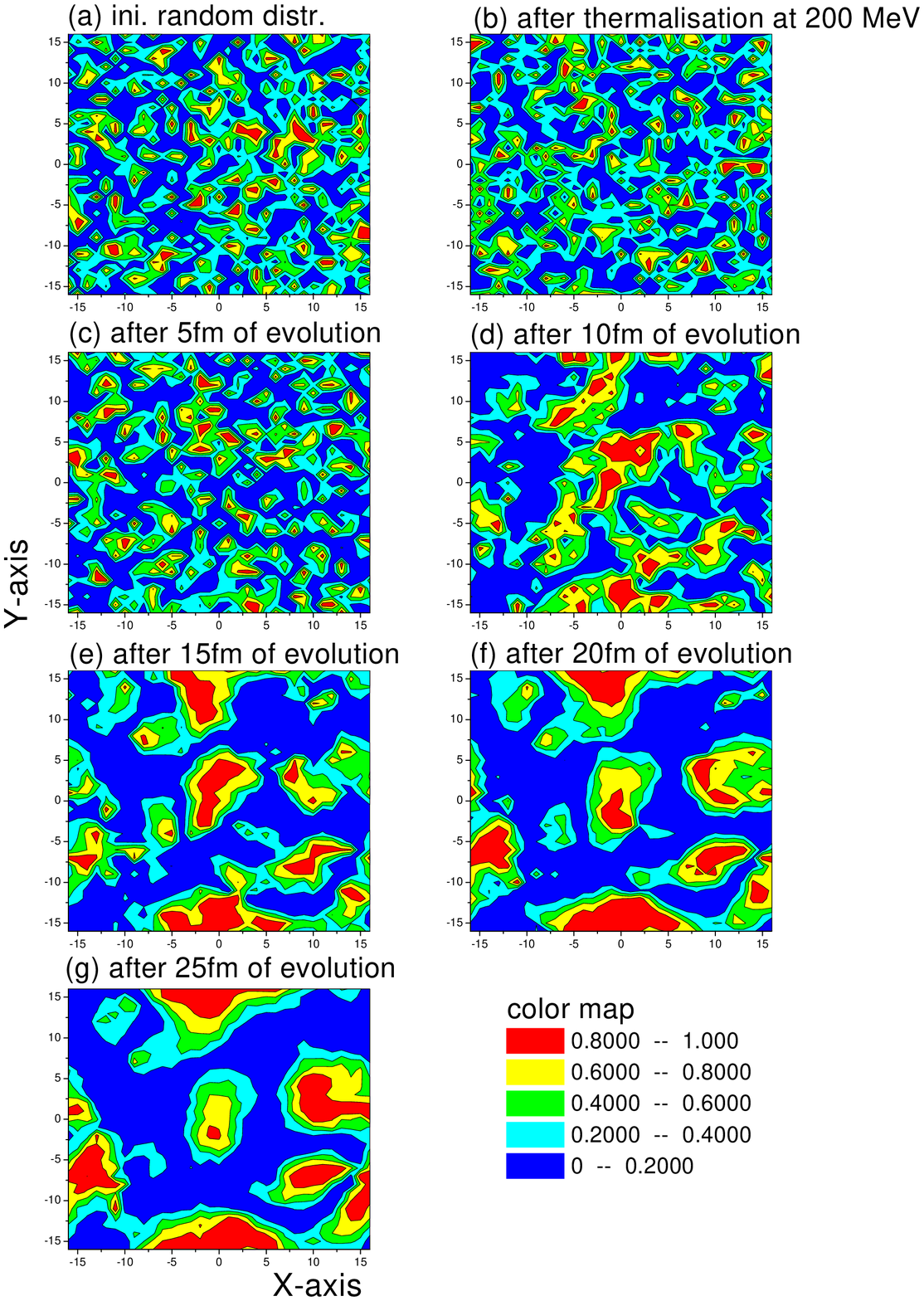,height=10cm,width=10cm}}
\caption{Same as fig.5 but for rapidity Y=4.}
\end{figure}

\begin{figure}
\centerline{\psfig{figure=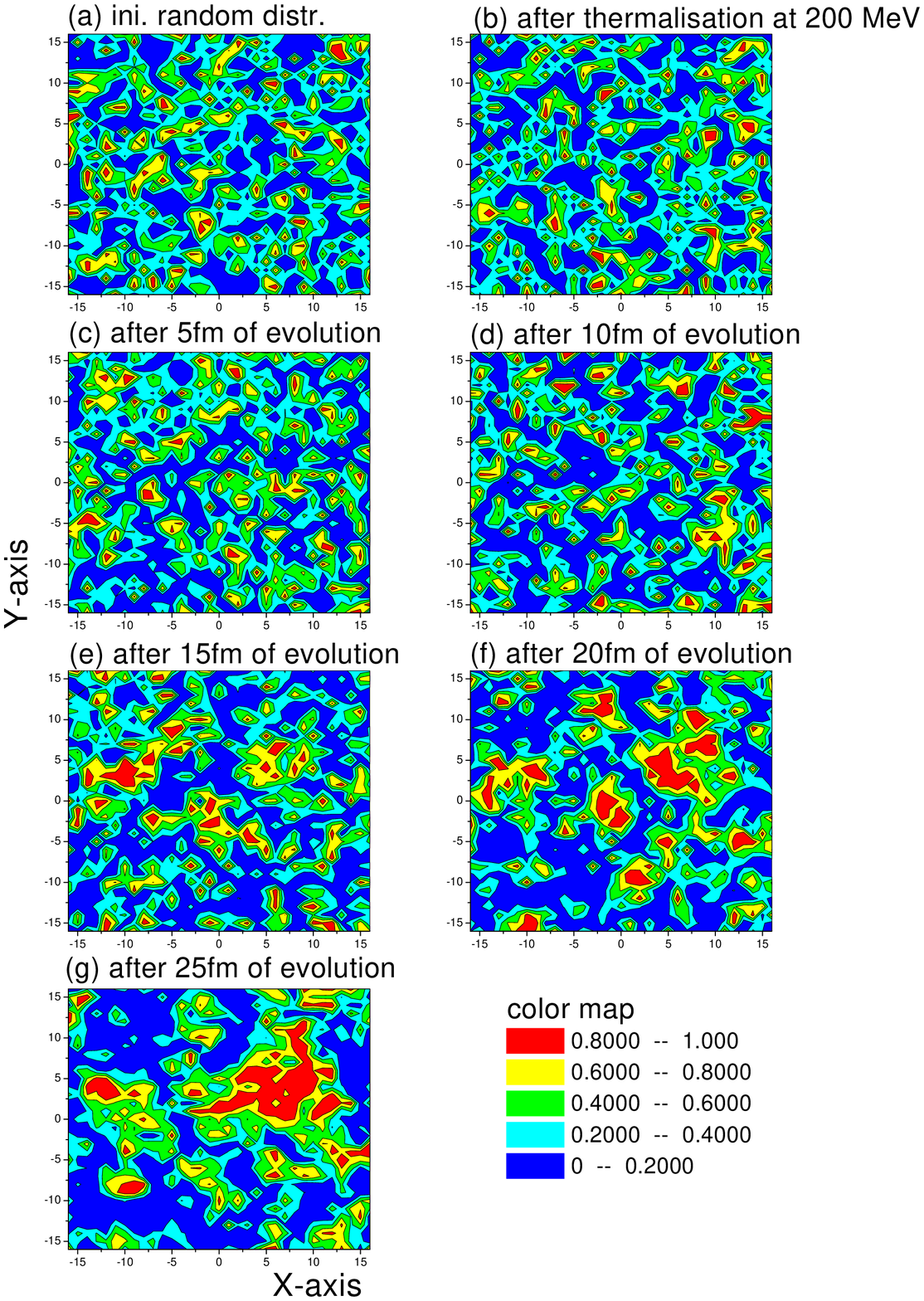,height=10cm,width=10cm}}
\caption{Contour  plot  of the neutral to total pion ratio, at
rapidity Y=0. The explicit symmetry breaking term is omitted i.e.
pions are massless. The cooling law corresponds to  slow  cooling
law.  Different  panels  shows  the  evolution  of  the  ratio at
different times. Domain like structure is evident  at  very  late
times.}
\end{figure}

\begin{figure}
\centerline{\psfig{figure=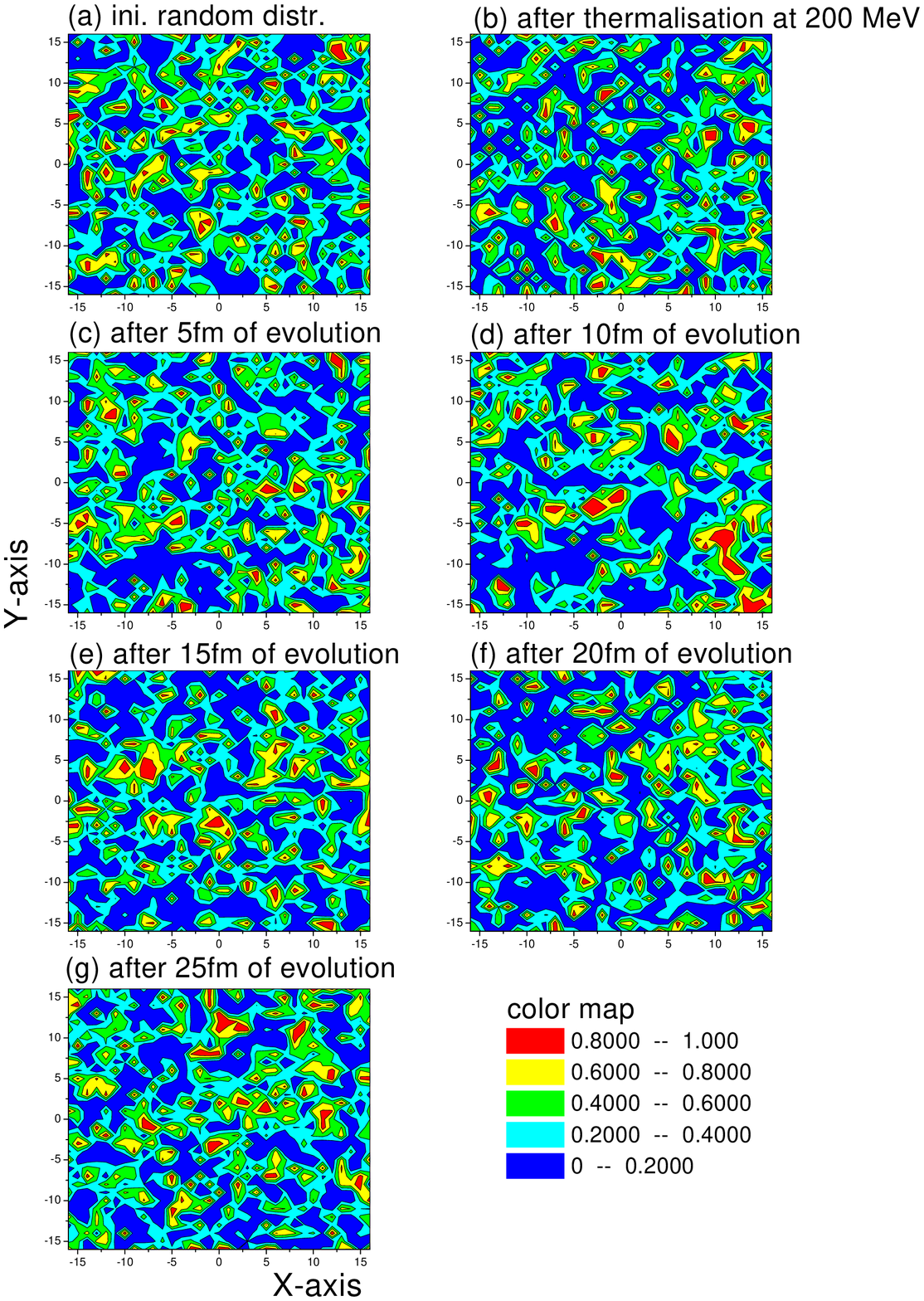,height=10cm,width=10cm}}
\caption{Contour  plot  of the neutral to total pion ratio, at
rapidity Y=0. The explicit symmetry  breaking  term  is  included
i.e.  pions  are  massive.  The  cooling  law corresponds to fast
cooling law. Different panels shows the evolution of the ratio at
different times. No Domain like structure is evident even at late
times.}
\end{figure}

\begin{figure}
\centerline{\psfig{figure=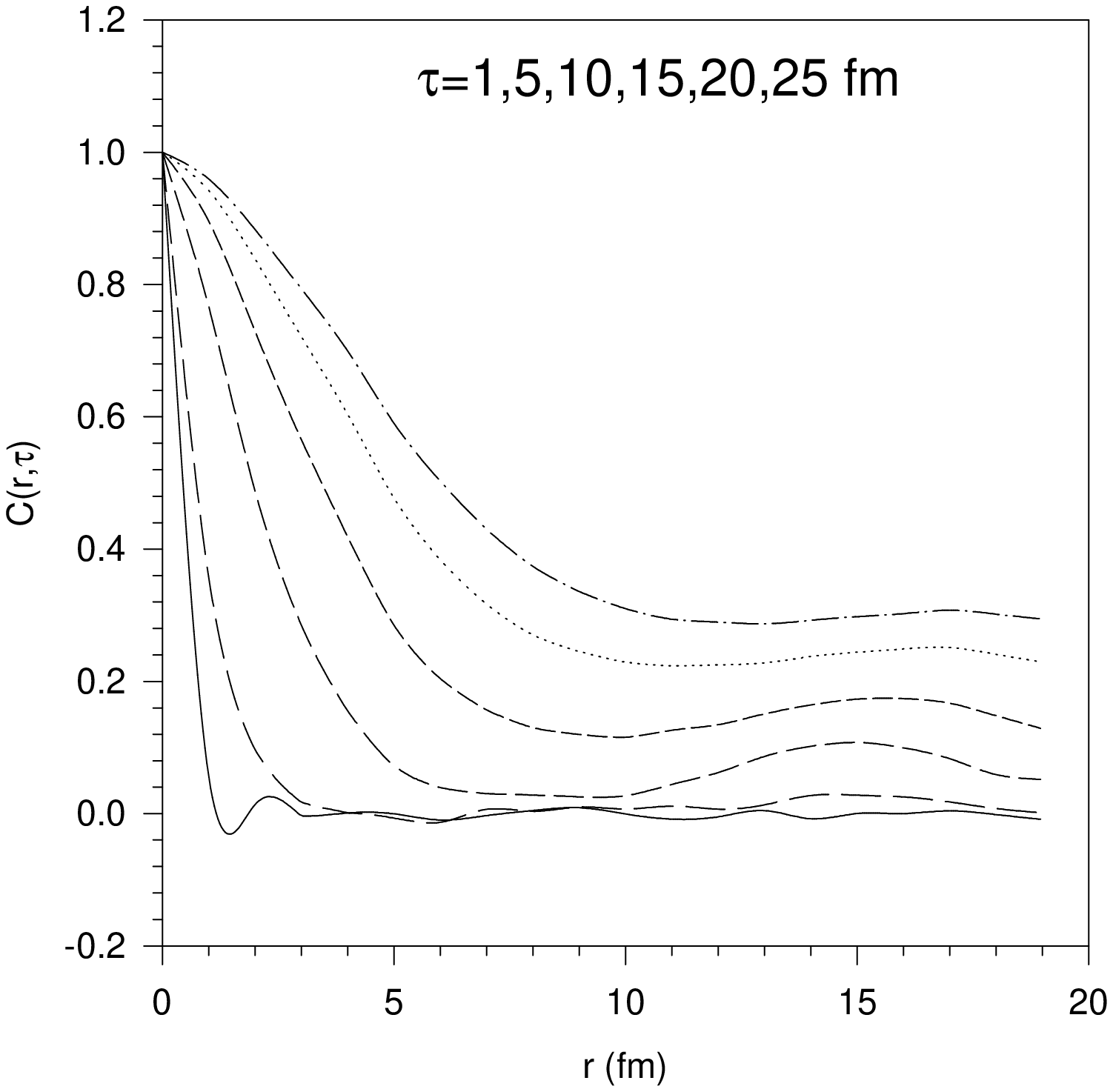,height=10cm,width=10cm}}
\caption{Evolution of the correlation function with time for zero
mass  pions  at  rapidity  Y=0.  Cooling  law corresponds to fast
cooling law. Long range  correlation  develops  at  late  times.}
\end{figure}

\begin{figure}
\centerline{\psfig{figure=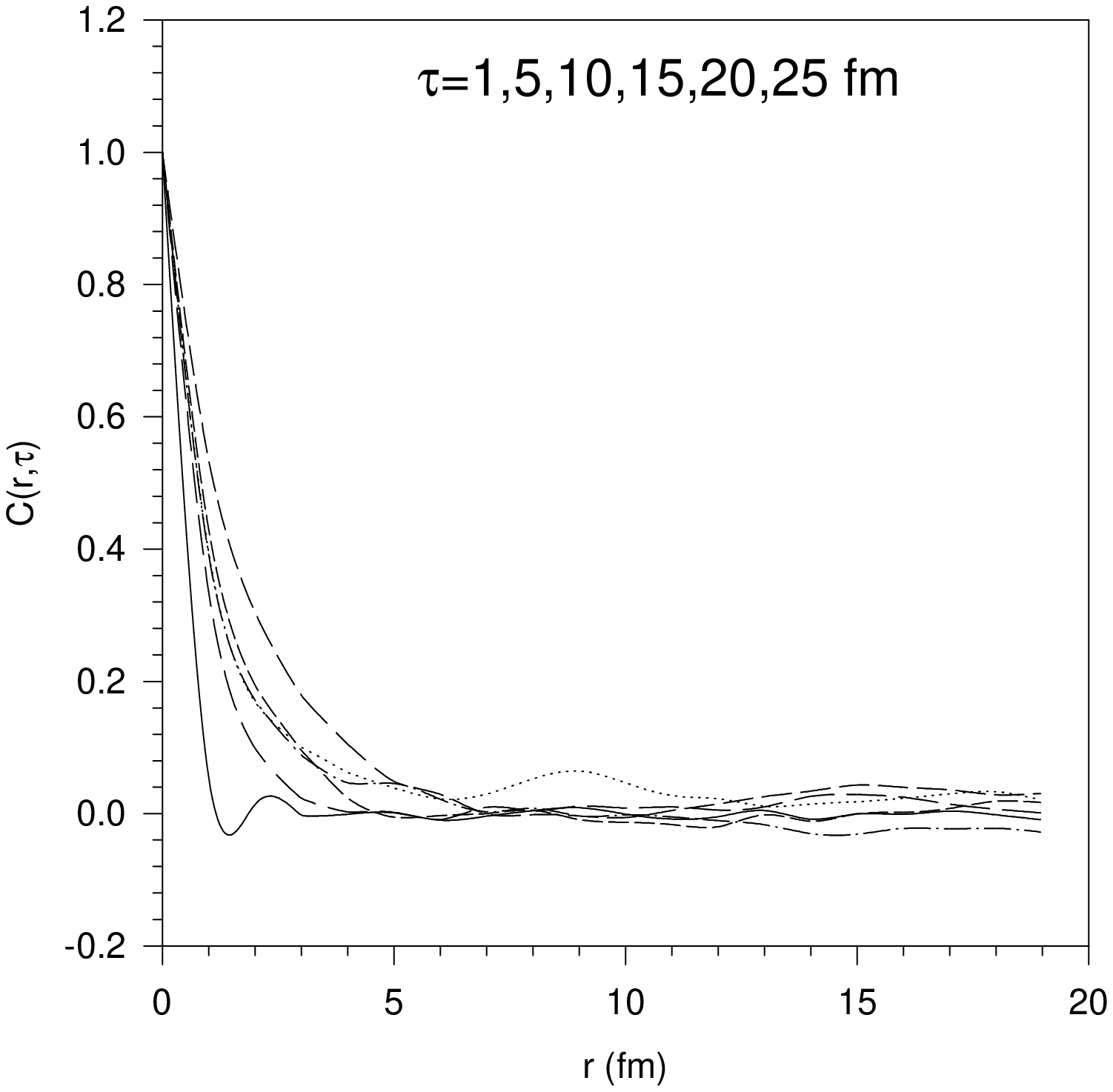,height=10cm,width=10cm}}
\caption{Evolution  of  the  correlation  function  with time for
massive pions at rapidity Y=0. The  cooling  law  corresponds  to
fast  cooling  law.  No  long  range correlation develops at late
times.}
\end{figure}

\begin{figure}
\centerline{\psfig{figure=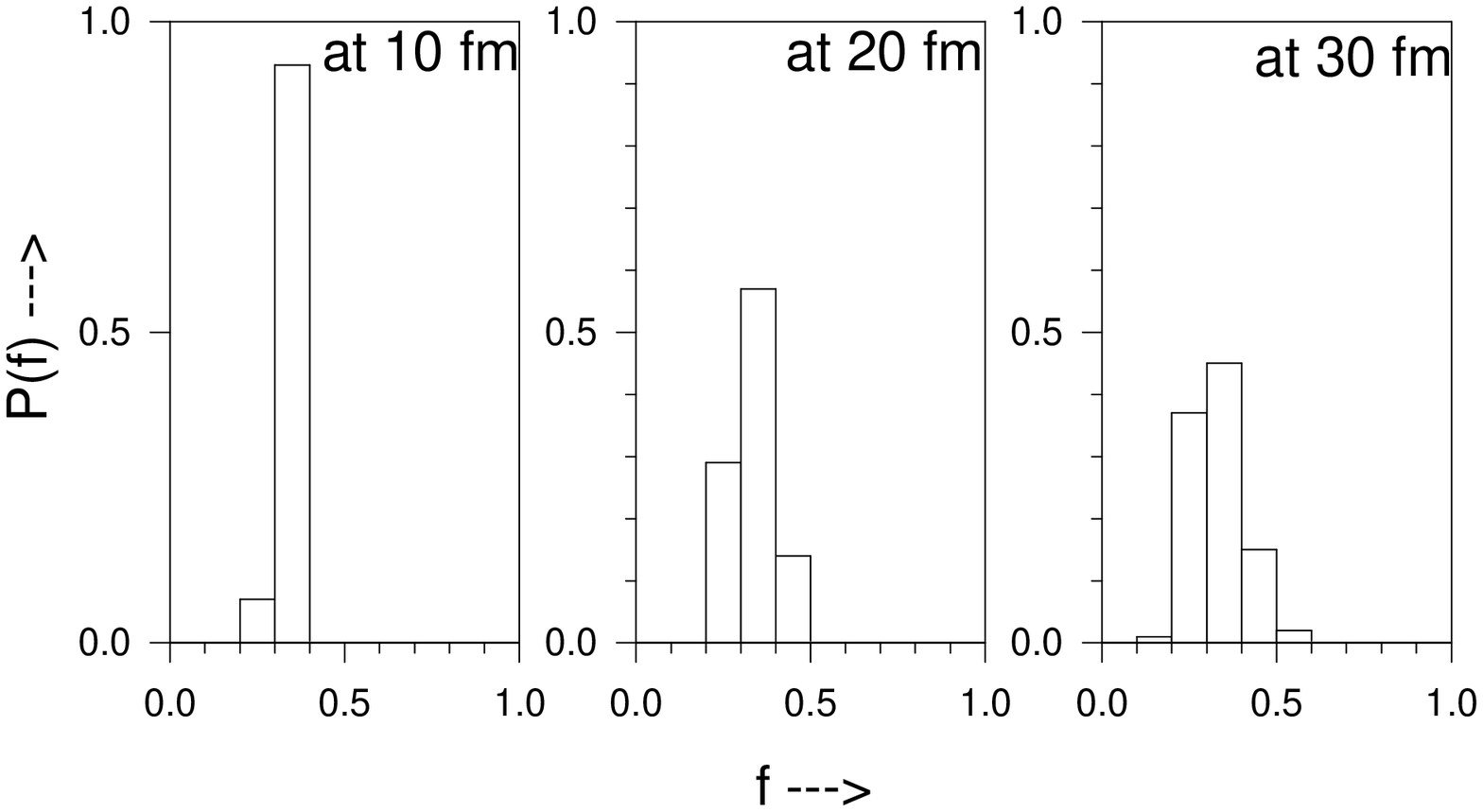,height=10cm,width=10cm}}
\caption{Probability distribution for the neutral to  total  pion
ratio, at different times. The symmetry breaking term is omitted.
The cooling law corresponds to fast cooling law.}
\end{figure}

\begin{figure}
\centerline{\psfig{figure=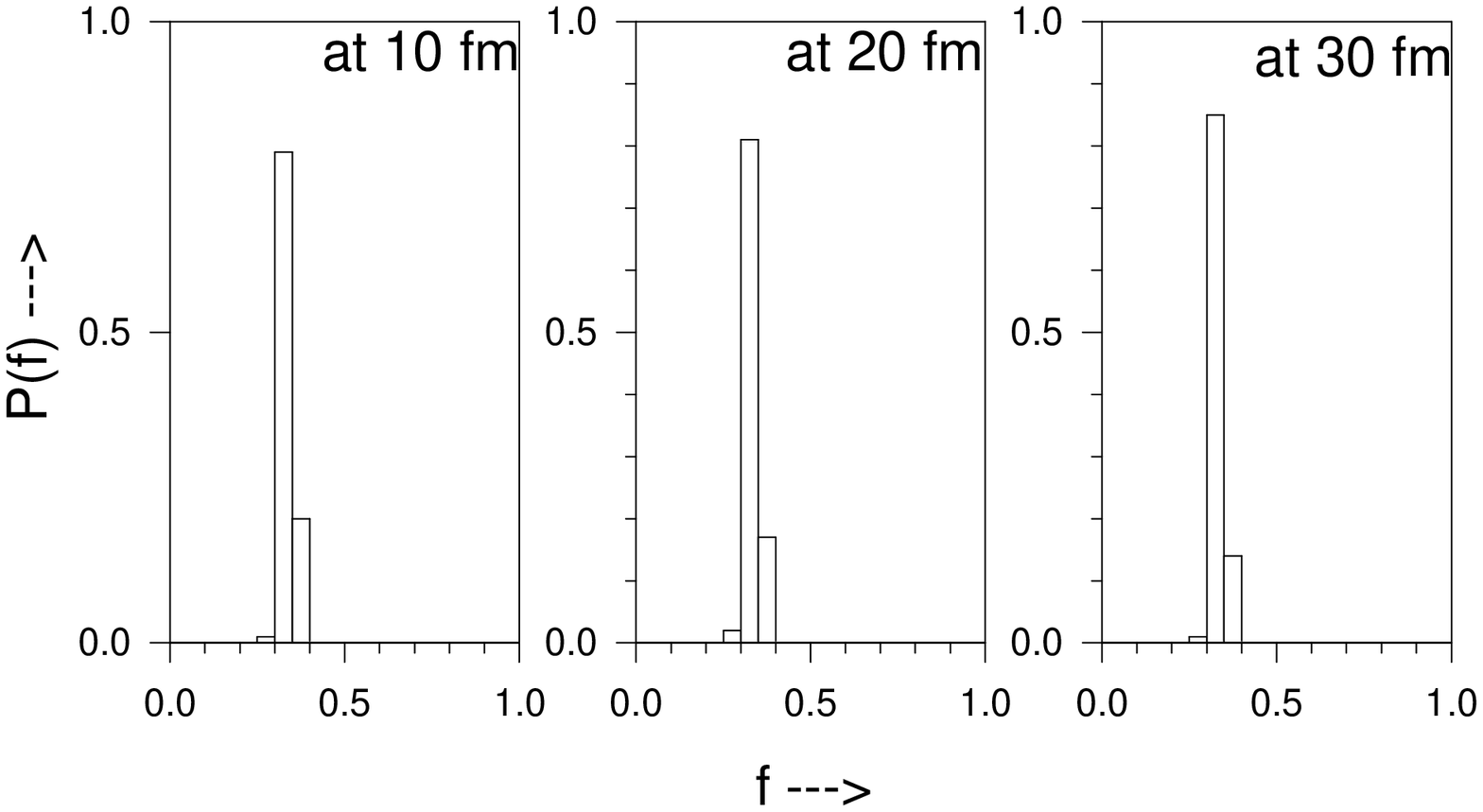,height=10cm,width=10cm}}
\caption{Probability distribution for the neutral to  total  pion
ratio, at different times. The symmetry breaking term is included.
The cooling law corresponds to fast cooling law.}
\end{figure}
\end{document}